\documentclass[twocolumn,unsortedaddress,floatfix]{revtex4-2}

\usepackage{siunitx}
\usepackage{graphicx}
\usepackage[colorlinks]{hyperref}
\usepackage{xcolor}

\definecolor{linkcolor}{RGB}{0,83,166}
\hypersetup{
  colorlinks = true,
  allcolors = {linkcolor}
}

\begin{document}

\title{Qubit spin ice}

\author{Andrew D. King$^{1,*}$, Cristiano Nisoli$^{2,**}$,  Edward D. Dahl$^{1,2}$, Gabriel Poulin-Lamarre$^{1}$, and Alejandro Lopez-Bezanilla$^{2}$}

\affiliation{$^1$D-Wave Systems, Burnaby, British Columbia, Canada, V5G 4M9, Canada}

\affiliation{$^2$Theoretical Division, Los Alamos National Laboratory, Los Alamos, NM, 87544, USA}

\email{$^*$aking@dwavesys.com, $^{**}$cristiano@lanl.gov}

\date{\today}

\begin{abstract}
  Artificial spin ices are frustrated spin systems that can be engineered, wherein fine tuning of geometry and topology has allowed the design and characterization of exotic emergent phenomena at the constituent level.  Here we report a realization of spin ice in a lattice of superconducting qubits.  Unlike conventional artificial spin ice, our system is disordered by both quantum and thermal fluctuations.  The ground state is classically described by the ice rule, and we achieve control over a fragile degeneracy point leading to a Coulomb phase.  The ability to pin individual spins allows us to demonstrate Gauss's law for emergent effective monopoles in two dimensions.  The demonstrated qubit control lays the groundwork for potential future study of topologically protected artificial quantum spin liquids.
\end{abstract}

\maketitle

\begin{figure}\includegraphics[width=8.6cm]{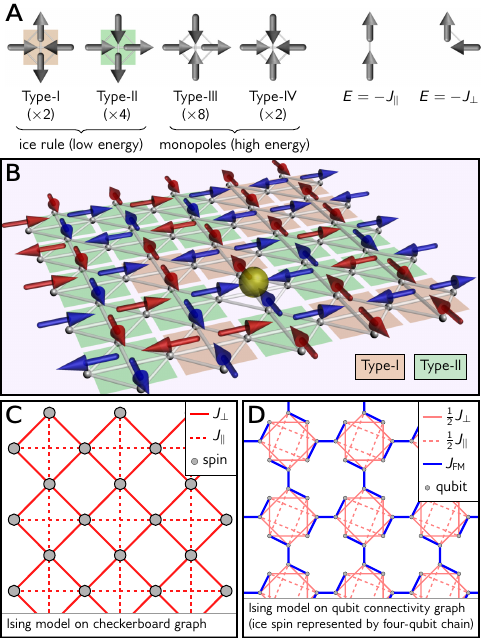}
  \caption{{\bf Realizing square ice in a quantum annealer}.  {\sffamily\bfseries \bf A}:  Each square ice vertex involves four dipoles that point in or out; six of sixteen configurations satisfy the two-in-two-out ice rule.  The remaining vertices have charge $\pm2$ (Type-III) or $\pm4$ (Type-IV), and host monopoles.  Two coupling energies $J_\parallel$ and $J_\perp$ determine energetic preference between Type-I and Type-II vertices in the artificial spin ice. {\bf B}: Schematic of a square spin ice with dipoles colored according to their Ising representation (red=$1$, blue=$-1$), with a monopole of  net charge of $+2$ (yellow sphere).   {\bf C}: Ising spin representation of {A}. Ice vertices (squares with dotted diagonals) form corner-sharing checkerboard plaquettes of four Ising spins (circles) each.  {\bf D}:  Embedding of C into the qubit connectivity graph (see Supplemental Material).  To realize the checkerboard geometry, each spin in C is represented using four qubits (circles), which are forced to act collectively by strong ferromagnetic coupling (blue lines).  Two chains impinging on the same ice vertex are coupled using two antiferromagnetic couplers, so each $J_\perp$ or $J_\parallel$ term is split into two equal coupling terms (red lines).}
  \end{figure}

\begin{figure*}\includegraphics[width=1 \linewidth]{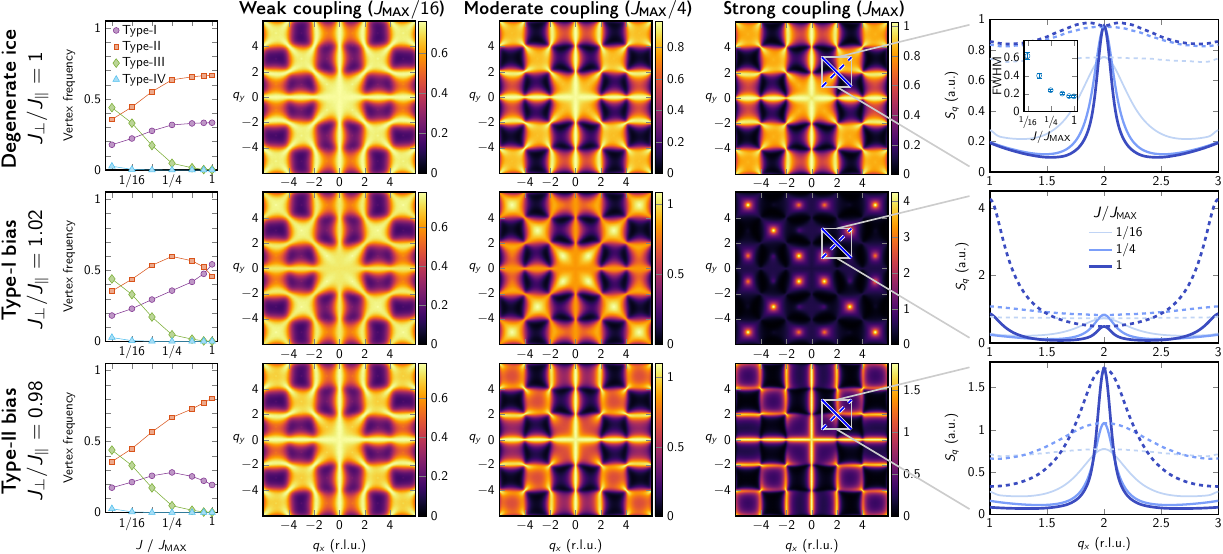}
  \caption{{\bf Experimental results: fine-tuning the ice ensemble.} From top row to bottom: the degenerate case $J_\perp=J_\parallel$, then $J_\perp=1.02J_\parallel$, and $J_\perp=0.98J_\parallel$. First column: frequencies of the different vertex types (Fig.~1-A) versus the energy scale $J/J_{\mathrm{MAX}}$, averaged over many measurements in a $14\times 14$ ice system. Second to Fourth: structure factor $S(\mathbf q)$ (arbitrary intensity units) for varying coupling energy scale in the three cases, in reciprocal lattice space.  Fifth column: cross sections of $S(\mathbf q)$ at the pinch points. The degenerate case shows the pinch point singularity associated with the Coulomb phase. The full width at half max (FWHM, inset) is the reciprocal correlation length, decaying as coupling energy increases (temperature is constant) and saturating in the strong coupling limit due to finite system size.  Tuning away from degeneracy results in the expected Bragg peaks (middle row) and collinear correlations (bottom row).}
  \end{figure*}

\begin{figure*}\includegraphics[width=10.2cm]{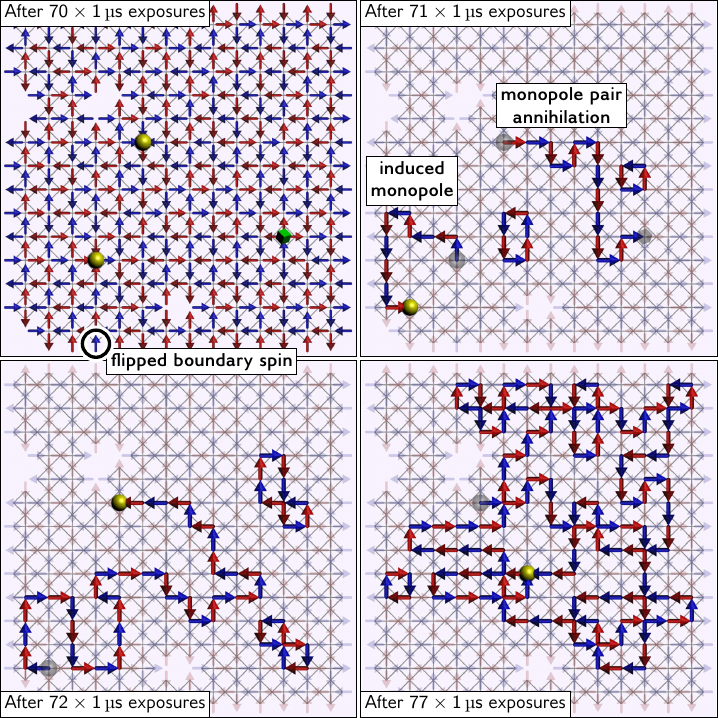}
  \caption{{\bf Gauss's law and monopole kinetics.} Successive QA samples with monopoles (positive ones denoted by yellow sphere, negative by green cube).  Boundary spins (including defect boundaries) can be clamped using per-qubit longitudinal fields.  Forcing antiferromagnetic alignment of boundary spins  gives no net flux of magnetization in the system; flipping a single boundary spin (shown) gives a net flux of $2$, and thus induces a monopole in the ground state.  The difference between two successive QA samples (faded arrows are unchanged from the previous spin state) shows the net motion of an induced monopole.  In this example we also see annihilation of two monopoles and small regions of local reconfiguration. Missing spins denote vacancies as described in text.}
  \end{figure*}

\begin{figure*}\includegraphics[width=\linewidth]{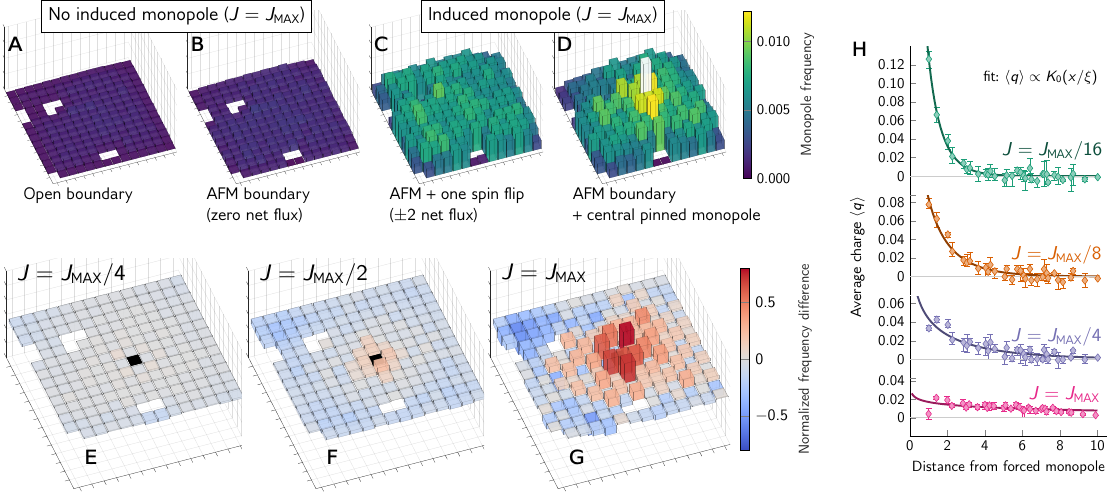}
\caption{{\bf Monopole frequency and interaction.}  Top row shows four boundary conditions: {\bf A}: Open boundary conditions; %, where no field $h$ is used;
  {\bf B}: Zero boundary flux, where boundary spins are forced to align antiferromagnetically (AFM); %with a Type-I background; 
  {\bf C}: Nonzero net flux, where a random boundary spin is flipped from the AFM background; {\bf D}: AFM boundaries with a central pinned monopole, where a central vertex is forced into a Type-III configuration.  Structural vacancies in the lattice are shown as missing data; in D the pinned monopole is represented by a white bar.  {\bf E--G}:  Using state C as a background density for an induced monopole, we take the difference between D and C, normalized by the average frequency in C, to demonstrate monopole-monopole entropic interaction;  {\bf H}: Entropic screening of charge shows good agreement with a Bessel decay form for weak coupling.  For strong coupling, the correlation length is large compared to finite system size, leading to a poor fit.  Data and error bars indicate the mean, maximum, and minimum values for a given Euclidean distance.}
 \end{figure*}

Artificial spin ices are systems of interacting components characterized by frustrated binary variables whose collective behavior emerges from local constraints based on the two-in-two-out ``ice rule'' (Fig.~1A).  They were initially introduced~\cite{Wang2006,tanaka2006magnetic} as analogues of the frustrated rare earth pyrochlores~\cite{Ramirez1999,Bramwell2001}, but then evolved  to generate, via frustration and disorder, exotic emergent phenomena not found in natural systems~\cite{nisoli2017deliberate}.   In simple ice-rule systems, the low-energy collective states can be described in terms of their excitations, which are emergent magnetic monopole quasiparticles~\cite{Castelnovo2008,ryzhkin2005magnetic2,perrin2016extensive,farhan2019emergent}. The most common artificial spin ice realizations have been based on lithographically patterned nanomagnets~\cite{Nisoli2013colloquium,skjaervo2019advances}. The more general set of ideas has been exported to other platforms, including colloids and superconducting vortices confined to bistable traps, and even to liquid crystals~\cite{ortiz2019colloquium} and exotic mechanics of soft modes~\cite{meeussen2020topological}. 

Here we implement a two-dimensional spin ice of superconducting qubits---macroscopic {\bf quantum} objects \cite{Harris2010a}---in a quantum annealing (QA) system, and drive it between low-energy quasi-classical states via primarily quantum rather than thermal fluctuations, thus opening the door to future studies of quantum phases in these systems~\cite{Henry2014, Chamon2019}.  Our ``qubit spin ice'' does not require fixed prefabrication; instead, its energetic coupling terms can be fine-tuned, and spins can be pinned at will.  We exploit this flexibility to demonstrate how Gauss's law emerges from geometric constraints in two dimensions. By fixing the total flux of magnetization into the system's boundary, we inject topological charges into the ground state, demonstrating induction of magnetic monopole quasiparticles which, unlike in dipolar implementations, interact purely entropically.

Square spin ice consists of a set of classical dipole spins placed along the edges of a square lattice, as shown in Fig.~1B.  The spins impinging on vertices realize sixteen different vertex configurations that are grouped by topology into Type-I, $\dots$, Type-IV (Fig.~1A).  The first two types obey the so-called ``ice rule'' (two spins point in, two point out) and are energetically favored in spin ice materials.  The other two violate the ice rule, as signalled by their topological charge (defined as the difference between spins pointing in and out) of $\pm 2$ and $\pm 4$ respectively, and are  monopole excitations.  Vertex energies are dictated by the antiferromagnetic couplings  $J_{||}$, $J_{\perp}$ between spins impinging on the vertex collinearly and perpendicularly respectively, and are: $\epsilon_{\mathrm{I}}=-4J_{\perp}+2J_{||}$, $\epsilon_{\mathrm{II}} = -2J_{||}$, $\epsilon_{\mathrm{III}}=0$, $\epsilon_{\mathrm{IV}}=4J_{\perp}+2J_{||}$.  The resulting system is geometrically frustrated.

Before exploring the phases of this ice system, we describe the QA system with which we realize square ice in its well-known Ising form \cite{Henry2014}.  The QA system comprises a set of superconducting flux qubits that interact via 2-body couplers \cite{Bunyk2014, sm}, physically realizing the transverse-field Ising model generically described by the Hamiltonian
\begin{equation}\label{eq:ham}
\mathcal H = \mathcal J \bigg( \sum_{\langle ij \rangle}J_{ij}\hat \sigma_i^z \hat\sigma_j^z + \sum_i h_i \hat \sigma_i^z\bigg) - \Gamma\sum_i  \hat \sigma_i^x,
\end{equation}
where $\hat \sigma_i$ are Pauli matrices describing the qubit degrees of freedom; the tensor $J_{ij}$ describes the action of the couplers and $h_i$ is a per-qubit longitudinal field.  The terms $J_{ij}$ and $h_i$ can be programmed at will; local fields $h_i$ are always set to zero except when specified.

Unlike in the Hamiltonians proposed to describe quantum spin ice in pyrochlores~\cite{gingras2014quantum}, we have no quantum entanglement in the 2-body coupling terms. Thus, in absence of the transverse field $\Gamma$, the ground state of $\mathcal H$ is a set of Fock states that can be mapped into purely classical ones, namely the Fock product of eigenvectors of the Pauli matrices $\hat \sigma^z$. However, switching on the transverse field entangles the binary quantum variables, subjecting them to quantum fluctuations.  

At finite temperature, the QA system relaxes toward the thermal equilibrium of $\mathcal H$.  Rapidly decreasing $\Gamma$ and increasing $\mathcal J$ projects this thermal distribution to the $\hat\sigma^z$ basis.  This approach has recently been used for a variety of quantum Ising systems at finite temperature~\cite{Harris2018,King2018,Weinberg2019}, and has provided the most direct observation of many-body quantum annealing dynamics to date~\cite{King2019}.

To realize square ice in the QA system, we start with an Ising model \cite{Henry2014}:  We consider an alternating $A/B$ bipartition of vertices in the square lattice, calling a spin $S=1$ (red in Fig.~1B) if it points toward an $A$ vertex, or $S=-1$ (blue in Fig.~1B) if it points toward a $B$ vertex. This gives an antiferromagnetic Ising model on a checkerboard lattice (Fig.~1C) whose quantum extension is captured by the Hamiltonian in Eq.~1.  However, the geometry of qubit pairs that can be directly coupled is described by a ``Chimera'' graph \cite{Bunyk2014,sm}, which does not contain the required checkerboard lattice as a subgraph.  We therefore represent each ice spin with a chain of four qubits, forced to act collectively with strong ferromagnetic couplings.  These chains are intercoupled in a checkerboard geometry as shown in Fig.~1D, whose tiling across the QA chip gives a $14\times 14$ grid of ice vertices with eight site vacancies resulting from inoperable qubits.

We calibrate the system to degeneracy \cite{sm}, and use this point to define the nominal ratio $J_\parallel/J_{\perp}=1$.  The overall energy scale $J=J_\parallel=J_{\perp}$ is taken as the average total coupling between coupled four-qubit chains.  $J_{\mathrm{MAX}}$ indicates the maximum achievable value of this Ising energy scale.  Recall that the relationship between $J_\perp$ and $J_\parallel$ determines the relative energies of Type-I and Type-II vertices.  Three cases are possible.

When $J_{\perp}= J_\parallel=J$, the six ice-rule-obeying vertices (Type-I and Type-II)  have the lowest energy and the ground state is a degenerate manifold with residual entropy described by the degenerate six-vertex model~\cite{lieb1967residual}. Its elementary excitations are monopoles (Type-III), and the crossover temperature into the ice state is $T_{\mathrm{ice}} \simeq J$, or half the energy of a monopole.

When $J_{\perp}>J_\parallel$ the Type-I vertices have the lowest energy  and their tiling forms a long-range ordered, classical ground state typical of the early antiferromagnetic artificial spin ice realizations~\cite{Wang2006,Morgan2011}. 

When  $J_{\perp}<J_\parallel$ the Type-II vertices have the lowest energy and their tiling forms a disordered ground state (``line state'') of sub-extensive entropy \cite{perrin2016extensive}.

To probe equilibrium properties in these cases we strobe the transverse field in a repeated relaxation protocol \cite{King2019}, starting from a randomly generated classical spin state.  We simultaneously turn on both quantum ($\Gamma\approx J_{\mathrm{MAX}}/3$) and thermal ($T =10 mK \approx J_{\mathrm{MAX}}/12$) fluctuations for an exposure time of $256\ \mu s$, then turn them off rapidly and read out a projected classical spin state.  A single programming of the chip repeats this on/off cycle of fluctuations 128 times, producing a chain of 128 classical spin states---each is in or near the ground state manifold of the classical Ising Hamiltonian.  We perform many repeated experiments with $J$ ranging from $J_{\mathrm{MAX}}/32$ to $J_{\mathrm{MAX}}$.  Classical spin states for the Ising model depicted in Fig.~1D are read from the QA, then converted to states for the checkerboard Ising model (Fig.~1C), which in turn are converted to dipole configurations (Fig.~1B, Fig.~3).  See \cite{sm} Fig.~S7 for an example.

Figure~2 shows the results of this measurement for the $14\times 14$ ice system at different energy scales $J$ for the three cases $J_\perp = J_\parallel$, $J_\perp > J_\parallel$, and $J_\perp < J_\parallel$.  The first column shows the relative frequency of vertex types throughout the lattice; the other columns show the spin structure factor, including pinch-point detail, computed from average spatial correlations \cite{sm}.  A larger coupling $J$ leads to ensembles closer to the predicted ground states, with monopole excitations appearing only rarely. At the degeneracy point $J_\parallel=J_{\perp}$ the relative occurrences of Type-I and Type-II at high $J$ very closely match those expected in the monopole-free degenerate ground state of spin ice, described by the six-vertex model~\cite{lieb1967residual}. From the above, the coupling at the crossover to the ice manifold can be estimated to be $J_{\mathrm{ice}}\simeq J_{\mathrm{MAX}}/12$ and thus the data points $J/J_{\mathrm{MAX}}=1/16,1/8$ sit near the crossover, where monopoles become sparse.

Minimal ($2\%$) tuning of $J_{\perp}/J_\parallel$ away from degeneracy leads to the relative promotion of Type-I or Type-II vertices, with an effect that increases with $J$.  The strong sensitivity to degeneracy lifting is also apparent in the static spin structure factor $S(\mathbf q)$. For weak coupling, bias toward Type-I or Type-II is barely perceptible.  For strong coupling we see at degeneracy the familiar, transverse structure---including the typical pinch-point singularities---associated with the Coulomb phase in an algebraic spin liquid~\cite{henley2010coulomb}.  As the system is tilted toward Type-I vertices, the structure factor becomes dominated by the Bragg peaks of the long-range N\'eel ordering.  Likewise tilting the system away from Type-I vertices shows a line-state of Type-II vertices with long-range collinear spin-spin correlation.

We now concentrate on the degenerate case $J_{\perp}=J_\parallel$ .  Unlike in nano-magnetic realizations, longitudinal fields in qubit ice ($h_i$ in Eq.~1) can act on individual qubits.  We use these fields to pin a subset of spins and demonstrate induction of a single, itinerant monopole, by Gauss's law, as well as entropic interactions between monopoles. 

We first pin the boundary spins into a fixed antiferromagnetic boundary condition. By Gauss's law, the net flux of magnetization into the system is equal to the charge inside the system.  Therefore when we anneal the degenerate system under these boundary conditions we typically find zero monopoles, i.e., a ground state. But if we flip one fixed boundary spin, as shown in real-space in Fig.~3, we force a net flux, e.g. $=2$, into the system. Upon annealing, we then observe a net charge in the bulk, in the form of a free monopole of charge 2---in this case, the ground state contains a monopole by Gauss's law. Thus an isolated monopole charge is induced in absence of a corresponding anti-charge, by forcing a net flux on the boundaries.  Note that we must also pin the interior boundary spins produced by any vacancies caused by inoperable qubits.

We can also observe a quantum-activated random walk of these monopoles.  Thermal fluctuations have been used to drive spin dynamics in superparamagnetic nanoislands~\cite{Porro2013,farhan2013direct,kapaklis2014thermal}; here the spin dynamics are driven by both quantum and thermal fluctuations (Fig.~S5) \cite{sm}.  Modifying the previous protocol (with $J_\parallel=J_\perp=J_{\mathrm{MAX}}$), we strobe fluctuations for exposures of ${1}{\mu s}$, the minimum interval permitted by the control circuitry.

For large $\Gamma$ or long exposure, one  expects that the quantum fluctuations would erase the system memory. Remarkably, however, for a carefully chosen value of $\Gamma=0.34 J_{\mathrm{MAX}}$ for ${1}{\mu s}$, the quantum-activated  system preserves memory of its previous classical state, and the qubit kinetics, while activated primarily by quantum fluctuations, reveals  monopole motion, monopole pair creation/annihilation, and collective flipping of closed loops of spins.

Figure~3 shows these phenomena in a sequence of samples from a QA experiment.  In the figure we highlight the difference between successive QA states. These are suggestive of a random walk of a monopole, although we cannot rule out intermediate creation and annihilation of additional monopole pairs.  Most samples contain only the isolated monopole induced into the ground state via a nonzero-flux boundary condition, but after $70$ exposures to fluctuations, a surplus monopole pair appears, making the ensuing sequence particularly interesting to visualize.  After $70$ exposures there are three monopoles for an overall net charge  $+2$, which matches the boundary flux. At $t={71}{\mu s}$ two monopoles of opposite charge have mutually annihilated, returning the system to the ground state. At $t={72}{\mu s}$ the induced monopole has moved again and by $t={77}{\mu s}$, several steps later, it has traversed much of the available space.  These timescales are in sharp contrast to the multi-second relaxation observations in nanoisland and colloidal implementations~\cite{farhan2013direct,farhan2019emergent,ortiz2016engineering}.  Example state sequences are shown in Movie S1 to S6 \cite{sm}.

Unlike in fully dipolar spin ice~\cite{Castelnovo2008}, our monopoles {\bf cannot} interact directly, because no appreciable long-range dipolar interaction exists between the qubits.  Monopoles, however, can be thought of as emergent quasiparticles in an underlying spin structure, and are therefore correlated by the divergence-free spin vacuum.  This correlation can be described as a pairwise interaction by which oppositely charged particles attract, but the attraction is merely a result of the degeneracy of spin configurations that are compatible with the monopole positions: it is an {\bf entropic} interaction and indeed its coupling constant depends on temperature~\cite{nisoli2020field}.  In this two-dimensional system it corresponds to the two-dimensional Coulomb law between charges $q_1$ and $q_2$ at distance $x$, which is logarithmic $\sim q_1 q_2 T\ln(x)$, and thus leads to a Bessel screening, or  $\langle q(x)\rangle \propto K_0(x/\xi)$ where $K_0$ is the modified Bessel function, and $\xi$ is a temperature-dependent correlation length~\cite{nisoli2020field}.

We can probe this purely entropic screening between monopoles by pinning a monopole at the center of our geometry. Figure~4 demonstrates the result of this pinning and compares it to the boundary conditions we have described above.  With open boundaries or with zero net flux (Fig.~4, A and B), monopoles are absent in the ground state and therefore are only rarely observed after annealing. With a net flux of $2$ (as in Fig.~3), by Gauss's law a monopole is forced into the system's ground state.  Simulating the system repeatedly with random assignments of the flipped boundary spin, we find that the probability of finding a monopole is fairly flat across the lattice (Fig.~4C). Thus, when flux inside the system is fixed, the forced monopole is delocalized in the bulk, as one would expect.  In contrast, when a monopole is pinned at the center and boundaries enclose zero flux (Fig.~4D), we observe a second, free monopole in the bulk that cancels the net charge of the pinned monopole.  We see that at high coupling, the probability of finding a free monopole is maximal close to the pinned one.  We also observe evidence of an alternating decay pattern that depends on the parity of the grid distance ($\ell_1$ norm) between two vertices~\cite{nisoli2020field}.  This phenomenon is more visible in absence of site vacancies \cite{sm} (Fig.~S6).

At lower coupling, more monopoles are available to screen the pinned charge, and  the {\bf relative} influence on monopole population is smaller (Fig.~4E--G). In Fig.~4H we plot average vertex charge, over many sampled states, as a function of lattice distance from the pinned monopole, revealing charge screening: for weak coupling, screening is indeed highly localized and in good agreement with the theoretically expected $K_0(x/\xi)$ Bessel decay form. The correlation length is known to be infinite in the ground state and thus grows with coupling.  For strong coupling the correlation length is comparable to system size (see also the plot of the FWHM in Fig~2, inset) and the screening becomes flatter.

Finally, in a square spin ice, the only kinetics that does not require an activation energy proceeds either via monopole motion or the much less likely collective flipping of entire loops of spins.  Not surprisingly we find that forcing itinerant monopoles into the sample, either via Gauss's law or pinned monopoles, leads to faster equilibration \cite{sm} (Fig.~S5) as the extra charge acts as a mobile catalyst for mixing.

The reconfigurability of our system will enable its generalization to a variety of lattice geometries~\cite{morrison2013unhappy}, including Kagome and 3D pyrochlore lattices~\cite{chern2014realizing} in near-term QA systems.  Despite evidence for a quantum-driven local dynamics, the observation of a macroscopic quantum phase~\cite{Henry2014, Chamon2019} remains out of reach at the operating temperature in this simulation \cite{sm}; however, these experiments offer vistas toward engineering a quantum spin liquid~\cite{Henry2014,Chamon2019,gingras2014quantum,stern2019quantum}.  Future quantum annealers will allow logical spins with greater low-temperature tunneling, expanding the study of collective quantum phenomena in frustrated systems.

%\clearpage

\subsection*{Acknowledgments}

We acknowledge the contributions of the processor development and fabrication teams at \mbox{D-Wave}, without whom this work would not be possible.  AK wishes to thank Arnab Banerjee (Purdue) and CN wishes to thank Peter Schiffer (Yale), Alan Bishop (LANL), and Nicolas Rougemaille (Institut N\'eel) for in-depth discussions.

\subsection*{Funding}

The work of ALB, EDD and CN was carried out under the auspices of the U.S.~DoE through the Los Alamos National Laboratory, operated by Triad National Security, LLC (Contract No. 892333218NCA000001). We thank ISTI at Los Alamos National Laboratory for financial support; work was also supported by the Institute for Materials Science (IMS) at Los Alamos under the program of ``IMS Rapid Response''. CN work was supported by a DOE LDRD grant.  The work of ADK, EDD, and GPL was funded by D-Wave.

\subsection*{Author contributions}

ALB, EDD and CN conceived the project.  CN, ADK, ALB, and EDD contributed to experimental design.  EDD realized the embedding.  ADK performed the QA experiments and data analysis.  GPL calibrated the QA processor and performed supporting measurements.  CN performed theoretical analysis and drafted the manuscript with ADK.  All authors contributed to the final version of the manuscript.

\subsection*{Competing interests}

ADK, EDD, and GPL are current or recent employees of and/or holders of stock options in D-Wave, which designs, manufactures, and sells the QA apparatus used in this work, and declare a competing interest on that basis.  CN and ALB declare no competing interests.

\subsection*{Data and materials availability}

Experimental data are available online \cite{data}.

%Here you should list the contents of your Supplementary Materials -- below is an example. 
%You should include a list of Supplementary figures, Tables, and any references that appear only in the SM. 
%Note that the reference numbering continues from the main text to the SM.
% In the example below, Refs. 4-10 were cited only in the SM.     

\clearpage

\onecolumngrid
\appendix

\section*{Supplementary materials}

\twocolumngrid

\section*{Materials and Methods}

\subsection*{QA methods}

The QA system used in this study was a D-Wave 2000Q system that uses radio-frequency superconducting quantum interference device (rf-SQUID) flux qubits to realize controllable spin-1/2 particles in a transverse-field Ising model.  The qubits are described in \cite{Harris2010a,Johnson2011}; the architecture is described in \cite{Bunyk2014}.  These qubits realize two $\sigma^z$-basis states $\mid\uparrow\rangle$ and $\mid\downarrow\rangle$ as a current circulating around the superconducting qubit body in one direction or the other (see schematic Fig.~1 in \cite{Johnson2011}).  The ability of these qubits to accurately realize a transverse-field Ising model is confirmed below, in ``Estimation of effective Ising Hamiltonian''. In this particular system, 2041 of the 2048 fabricated qubits and 5974 of 6016 couplers were operational.  The qubit connectivity lattice is a ``Chimera'' graph consisting of a $16\times 16$ grid of eight-qubit unit cells~\cite{Bunyk2014}.

To implement the checkerboard Ising lattice used for our artificial square ice, we use strong FM couplings to bind together four-qubit chains; each chain then represents a single logical spin of the checkerboard lattice as depicted in Fig.~1 (see also Fig.~S1 for embeddings of the $14\times 14$ lattice of checkerboard plaquettes, and a diagram drawn over a photograph of the chip).  The available programmable range for each coupler is $[-2,1]$ (in units of Ising energy scale $\mathcal J(s)$, as discussed later).  Negative values indicate FM coupling, and in a four-qubit chain we program all three couplers to $-2\mathcal J$.

The  $J_\perp$ and $J_\parallel$ couplers within an ice vertex are implemented using AFM couplers in a Chimera unit cell; each $J_\perp$ and $J_\parallel$ term is implemented using two couplers in the QA system (Fig.~S2A).  The difference between these two coupling geometries leads to a slight lifting of degeneracy between Type-I and Type-II configurations.  In other words, if we program $J_\perp$ and $J_\parallel$ to be nominally equal, whether in experiment or in simulation, the effective inter-chain coupling due to $J_\perp$ will be stronger than the coupling due to $J_\parallel$.  The same phenomenon is discussed and compensated in Ref.~\cite{Harris2018}.  We compensate for the lifting of degeneracy by tuning the AFM couplers individually by between one and four percent, in the process of calibration refinement discussed below.

At $J=J_{\mathrm{MAX}}=1.92\mathcal J$ we set each of these AFM couplers to be $0.96$ before the calibration refinement, to leave space within the $[-2,1]$ coupling range for fine tuning.  This allows a total AFM coupling between two four-qubit chains of $J_{\mathrm{MAX}}=1.92\mathcal J$, spread across two couplers.

With this energy scale in mind, AFM couplers are programmed with an exchange strength depending on (1) the desired overall coupling strength $J/J_{\mathrm{MAX}}$, (2) bias between $J_\parallel$ and $J_\perp$, and (3) fine-tuning of the calibration as detailed below.

Since the four-qubit chains impinging on an ice vertex extend by one unit cell in each direction from the unit cell containing the in-vertex coupling terms (Fig.~1), the $16\times 16$ grid of unit cells gives us a $14\times 14$ grid of ice vertices.  In the presence of a defective qubit or coupler, we modify the problem by first removing a set of four-qubit chains so that no remaining chain contains or is incident to a defective device, and second by removing (by setting to zero) all remaining AFM coupling terms within each ice vertex that has fewer than four of its chains remaining.  The resulting lattice is shown in Fig.~3.

The QA system has a small amount of unwanted disorder in the effective coupling terms.  This disorder arises from various sources including device variation and crosstalk.  To suppress this disorder, we average our experiments over multiple realizations (embeddings into the qubit connectivity graph).  Each experiment uses 20 randomly generated embeddings of the simulated Ising model (with the same set of vacancies; two are shown in Fig.~S1).  Thus we run experiments using different sets of qubits and couplers to implement the same system, and present average statistics.

Fig.~S1C shows a photograph of a portion of the QA processor with details of the spin ice representation drawn over.  Since the qubit bodies are elongated wire loops, they appear as line segments, and the couplers appear as circles, opposite to the representations in Fig.~1 and Fig.~S1A--B.  The four colours correspond to the four 4-qubit chains representing the four ice spins impinging on a single ice vertex; the ice vertex naturally corresponds to a plaquette in the checkerboard lattice.  In Fig.~S1D the ferromagnetic couplers are represented by colored dots.  Each $J_\perp$ and $J_\parallel$ coupling is physically realized by two couplers (white and black dots, respectively).

  Although the embeddings differ locally, they share the same general structure: a $14\times 14$ grid of ice vertices, whose internal AFM couplings are in the unit cells of the ``Chimera'' qubit topology.  Each ice vertex intersects with four four-qubit FM chains.  Both embeddings have the same pattern of vacant sites.  For each experiment we present average statistics over 20 randomly generated embeddings. 
\subsubsection*{Calibration refinement}

Within each embedding we refine the calibration of the qubits and couplers via iterative fine-tuning.  We run two types of experiment sets: varying the phase, as depicted in Fig.~2, and varying the boundary/monopole conditions, as depicted in Fig.~4.  Both sets include degenerate ice with open boundaries, and we use these specific observations to refine the calibration.

After each iteration of the experiment set, we gather statistics: observed magnetization of the qubits, and spin-spin correlation of antiferromagnetically-coupled qubit pairs.  We take these statistics for the degenerate ice setting with open boundaries, and make two assumptions: the magnetization of each qubit should be zero, and the spin-spin correlation of each antiferromagnetically-coupled qubit pair should be the same (this assumption compensates for the degeneracy lifting between $J_\perp$ and $J_\parallel$ terms described above).  And after each iteration, we make small adjustments to the Hamiltonian terms to enforce these assumptions, using a gradient descent method.  If two antiferromagnetically-coupled qubits are two highly correlated, we strengthen the coupling, and vice-versa.  The resulting coupling terms have a standard deviation on the order of $0.001$---an order of magnitude smaller than the $J_\perp/J_\parallel$ bias induced and studied in Fig.~2.  If a qubit has a nonzero observed magnetization, we tune a ``flux-bias offset'' to bias the qubit toward the zero-magnetization degeneracy point.  This type of refinement has emerged as an important ingredient in quantum annealing of degenerate systems \cite{King2018, King2019, Kairys2020}.

\subsection*{Reverse anneal chains}

Here we describe the annealing protocol by which our spin ice system is relaxed.  Each experiment described in this work consists of many repeated calls to the QA system.

The QA realizes the TFIM Hamiltonian (Eq.~1) in a parameterized form using an annealing parameter $0\leq s\leq 1$:
\begin{equation}\label{eq:qaham}
\mathcal H = \mathcal J(s) \bigg(\sum_{\langle ij \rangle}J_{ij}\hat \sigma_i^z \hat\sigma_j^z + \sum_i h_i \hat \sigma_i^z\bigg) - \Gamma(s)\sum_i  \hat \sigma_i^x.
\end{equation}
For each call, the QA system is programmed with the terms $J_{ij}$ and $h_i$, initialized in a random classical spin state, then evolved by modulation of $s$.  A chain of 128 classical output states is generated; between one output state and the next, quantum and thermal fluctuations are turned on (by reducing $s$ from $1$ to a value $s^*$ that gives the desired parameters $\mathcal J(s^*)$ and $\Gamma(s^*)$), held for a pause duration $t_p$, then turned off (by increasing $s$ from $s^*$ to $1$).  For Figs.~2, 4, and S6, $t_p= 256 \mu s $.  For Figs.~3 and S5, $t_p= 1 \mu s$.  The rate of change of $s$ is denoted $dt/ds = t_q$, which throughout this work is $ 1 \mu s$, the fastest allowed by the system.

This cycle of turning quantum and thermal fluctuations on and off is shown in Fig.~S3.  Between each reverse anneal is a readout taking $0.2 ms$ and a pause of $0.5 ms - t_p$; we denote this combined inter-anneal wait by $t_w = 0.7 ms - t_p$.  Note that quantum fluctuations are induced by the transverse field $\Gamma$, whereas thermal fluctuations arise from the system coupling to the environment; experiments are run at a fixed temperature of $T= 10$ mK.  Thus the strengths of the quantum and thermal fluctuations relative to the Ising energy scale are given by $\Gamma/J$ and $T/J$ respectively.

We repeat the 128-step QA chain many times for each experiment, reporting average statistics.  The main results (Figs.~2 and 4) reflect 200 repetitions using 20 lattice embeddings each; the first 16 steps of each 128-step chain are discarded as burn-in.  Thus each panel of Fig.~2 and Fig.~4 is derived from a total of $448,000$ samples (except for panels in the first column of Fig.~2, which present seven coupling energy scales using a total of $3,136,000$ samples).

\subsection*{Estimation of effective Ising Hamiltonian}

The rf-SQUID flux qubits have multiple energy levels and provide an imperfect approximation to spins in a transverse field Ising model.  We therefore follow methods set out in Ref.~\cite{King2019} to estimate the effective coupling and tunneling terms in the transverse-field Ising model (TFIM) Hamiltonians investigated.  In doing so we verify that the qubits used herein provide an accurate realization of Ising spins.  There are three relevant systems:
\begin{enumerate}
\item The QA system is made up of rf-SQUID flux qubits arranged in a Chimera topology (Fig.~S1), in which four-qubit chains are bound together using strong FM couplings.
\item The qubits provide an approximate physical realization of a TFIM Hamiltonian in the same Chimera topology (Fig.~1D), in which each degree of freedom is an ideal two-level spin-1/2 moment.
\item Finally, the Chimera TFIM is used to approximately realize the Ising square ice system on the checkerboard lattice (Fig.~1C), using each four-qubit FM chain of Chimera spins to represent a single logical ice spin.
\end{enumerate}
Here we describe the extraction of effective TFIM parameters from the physical qubit parameters.  Separate from this, the effective qubit temperature of $10$mK is measured as in \cite{Johnson2011} SM p.~8.

We follow the methods of Ref.~\cite{King2019} (SM Section 3) across a range of annealing parameters $s$.  The experiments described in the main body of this work were performed with $s=0.38$.  In the Chimera system, $J_\perp$ and $J_\parallel$ are realized with different geometry (Fig.~S2A), similar to the situation in \cite{Harris2018} (SM Fig.~S10).  Tuning of the $J_\perp = J_\parallel$ degeneracy point indicates that the effective difference is small (less than $3\%$) so for the extraction of TFIM Hamiltonian parameters we average out the two geometries.

The AFM and FM couplings in the gadget have strength $0.06J_{\mathrm{MAFM}}$ and $-2J_{\mathrm{MAFM}}$ respectively, where $J_{\mathrm{MAFM}}$ is the maximum AFM coupling between two flux qubits.  These values are chosen so that only chain-intact states (in which the strong FM couplers are respected), which are the most important to the experiment, contribute significantly to the low-energy spectrum.  We diagonalize the rf-SQUID flux qubit Hamiltonian of the 12-qubit gadget shown in Fig.~S2B) according to the independently measured qubit parameters (Fig.~S2D black lines).  Our aim is to determine TFIM energy scales $\mathcal J_{12}(s)$ and $\Gamma_{12}(s)$ such that the qubit Hamiltonian and TFIM Hamiltonian
\begin{equation}
\mathcal H_{12}(s) = \mathcal J_{12}(s)\bigg(\sum_{i,j}J_{ij}\sigma_i^z\sigma_j^z   + \sum_i h_i \hat \sigma_i^z \bigg) - \Gamma_{12}(s)\sum_{i}\sigma_i^x
\end{equation}
have similar eigenspectra.  Specifically we search for values for which the first seven eigengaps are almost identical (purple circles).  Using a best fit we extract $\mathcal J_{12}(s)$ and $\Gamma_{12}(s)$ (Fig.~S2E).  This gives the overall Ising energy scale $\mathcal J_{12}$ in of Hamiltonian $\mathcal H_{12}$; the total coupling energy between two four-qubit chains also requires us to consider the programmed terms $J_{ij}$.  To extract $\mathcal J_{12}$ we studied a small inter-chain coupling ($0.24J_{\mathrm{MAFM}}$ split across four coupling terms $J_{ij}$) to simplify the analysis of the eigenspectrum.  In the main experiments at maximum coupling, any two four-qubit chains intersecting the same ice vertex are coupled with a total inter-chain coupling of $1.92J_{\mathrm{MAFM}}$, split across two coupling terms $J_{ij}=0.96$.  The resulting maximum effective Ising energy scale of the inter-chain coupling is therefore $J_{\mathrm{MAX}}(s) = 1.92\mathcal J_{12}(s)$.  This is also shown in (Fig.~S2E).

We perform the same mapping to the three-spin TFIM Hamiltonian (Fig.~S2C)
\begin{equation}
\mathcal H_{3}(s) = \mathcal J_{3}(s)\sum_{i,j}J_{ij}\sigma_i^z\sigma_j^z + \Gamma_3(s)\sum_{i}\sigma_i^x,
\end{equation}
where each pair of spins is coupled with strength $0.24J_{\mathrm{MAFM}}$.  Again the eigengaps are shown to provide a good match in Fig.~S2D, and the extracted parameters are shown in Fig.~S2F.  As in the 12-qubit gadget, we note that the maximum effective inter-spin coupling energy is not $\mathcal J_3$ but rather $1.92\mathcal J_3$.

\subsection*{Magnetic structure factor}

The magnetic structure factor (Fig.~2) is defined as $S(\mathbf q) = \sum_{ij}(\delta_{ij}-q_iq_j)\langle\tilde S_i(\mathbf q)\tilde S_j(\mathbf q)\rangle$.  There, $\langle\tilde S_i(\mathbf q)\tilde S_j(\mathbf q)\rangle$ is the spin-spin correlation in reciprocal space, obtained by Fourier transform of the real-space correlation.  $S(\mathbf q)$ helps to determine the structure of a magnetic material, including the directions in which moments point in an ordered spin arrangement and the interaction between spins.  It plays an important role in the visualization of crystal structures because it relates to the intensity of a neutron beam reflection, which in turn depends on the magnetic structure giving rise to that reflection.

The images we obtain are the magnetic structure analogue of the image obtained in an optical microscope by recombination of the rays scattered by the object. In our case, the recombination of virtual diffracted neutron beams is mimicked by a mathematical calculation involving the Fourier transform of the square spin ice spin correlations. Magnetic structure factors were computed using the vector-based approach presented in Ref.~\cite{farhan2019emergent}.  This treatment computes the structure factor in analogy to neutron scattering experiments.  First, each Ising spin $s_i$ is converted into a dipole vector $\mathbf s_i$ as in the translation between the checkerboard Ising model and spin ice.  Then for each point $q$, we generate the intensity $S(\mathbf q)$ by measuring correlation perpendicular to the vector $\mathbf q$.  Defining $\mathbf s_i^\perp$ as the component of $\mathbf s_i$ perpendicular to $ \mathbf q$, and the vector $\mathbf r_{ij}$ as the vector from the site of $s_i$ to site of $s_j$, we define
\begin{equation}
  S(\mathbf q) = \frac 1N\sum_{i=1}^N\sum_{j=1}^N \mathbf s_i^\perp\cdot \mathbf s_j^\perp\exp(i\mathbf q\cdot\mathbf r_{ij}).
\end{equation}
This is in contrast to the real-space structure factor, for example used in Ref.~\cite{Henry2014}), which more simply computes
\begin{equation}
  S_{\mathrm{real}}(\mathbf q) = \frac 1N\sum_{i=1}^N\sum_{j=1}^N  s_is_j\exp(i\mathbf q\cdot\mathbf r_{ij}).
  \end{equation}
In Fig.~S4 we show the real-space structure factors $S_{\mathrm{real}}(\mathbf q)$ corresponding to the reciprocal-space structure factors in Fig.~2.  Note that while $S_{\mathrm{real}}(\mathbf q)$ is periodic on the Brillouin zone, $S(\mathbf q)$, by construction, is not.

\section*{Supplementary Text}

\subsection*{Effect of monopoles and quantum fluctuations on dynamics}

Under the conditions of the main experiments, with $s=0.38$ and $T\approx 10$ mK (measured as in \cite{Johnson2011} SM p.~8), the relevant energy scales are $T/J_{\mathrm{MAX}} = 0.083$, $\Gamma/J_{\mathrm{MAX}} = 0.34$ for the Chimera TFIM.  These parameters yield an approximate realization of the checkerboard Ising system with parameters $T/J_{\mathrm{MAX}} = 0.089$, $\Gamma/J_{\mathrm{MAX}} = 0.010$.  This indicates that the Ising coupling strength is not strongly affected by the embedding of logical ice spins into four-qubit chains, but the tunneling is suppressed by over an order of magnitude.
The ratio $T/\Gamma$ temperature in these experiments is therefore too high to reach the quantum Coulomb phase of the checkerboard Ising system (see \cite{Henry2014} Fig.~2).

Despite this, the quantum fluctuations accelerate dynamics in the system.  Fig.~S5 compares mixing rates for several values of $\Gamma/J$ with fixed $J/T$.  This is achieved by modulating the annealing parameter $s$ between $0.36$ and $0.41$ and tuning the programmable terms $J_{ij}$ and $h_i$ so that the classical part of $\mathcal H_{12}$, i.e.,
\begin{equation}
  \mathcal J_{12}(s)\bigg(\sum_{i,j}J_{ij}\sigma^z_i\sigma^z_j+\sum_ih_i\sigma^z_i\bigg)
\end{equation}
remains constant for each value of $s$.  As $\Gamma/J$ is increased, the sample-to-sample difference resulting from each exposure to fluctuations increases, with no accompanying increase in monopole count.  The data shown in Fig.~S5 correspond to the three closed-boundary configurations studied in Fig.~4, but show exposures of only $1 \mu s$, as in Fig.~3.

Additionally, the itinerant monopoles influence the mixing of the disordered ice system, since spins in the vicinity of a monopole can be flipped individually without changing the energy of the system.  This is not the case in the absence of a monopole, where either cotunneling or excitation is required to move the system away from its current state.  For exposures of $1 \mu s$ the boundary conditions have a significant impact: closed boundaries rely on excitations to drive mixing, evidenced by the fact that conditions C and D, with itinerant monopoles, mix considerably faster than condition B.  These boundary conditions also lead to lower populations of {\em surplus} monopoles, i.e., those that are not forced by the boundary condition (1 for condition C, 2 for condition D).  This is what one would expect: if a monopole pair appears spontaneously in the presence of an itinerant induced monopole, the result is, for example, a negatively charged monopole that can now annihilate with one of {\em two} positively charged monopoles rather than just one.

\subsection*{Vacancy-free lattices}

Fig.~S6 presents data analogous to Fig.~4, produced using a fully-yielded sublattice with no vacancies.  This allows us to suppress the effect of unwanted disorder and statistical noise by averaging over lattice symmetries (bottom half).

\subsection*{Conversion of QA output to square ice states}

Fig.~S7 shows the QA output sample corresponding to the upper-left panel of Fig.~3.  Four-qubit chains (blue lines) are contracted into logical variables by using an arbitrary spin as a representative.  Broken chains are rare enough that the method of parsing them is unimportant.  The qubit state is a state of the Ising model shown in Fig.~1D, and the checkerboard state is a state of the Ising model shown in Fig.~1C.  This particular embedding of the Hamiltonian onto the qubits is depicted in Fig.~S1C.

\clearpage
\onecolumngrid

\noindent    \includegraphics[height=19cm]{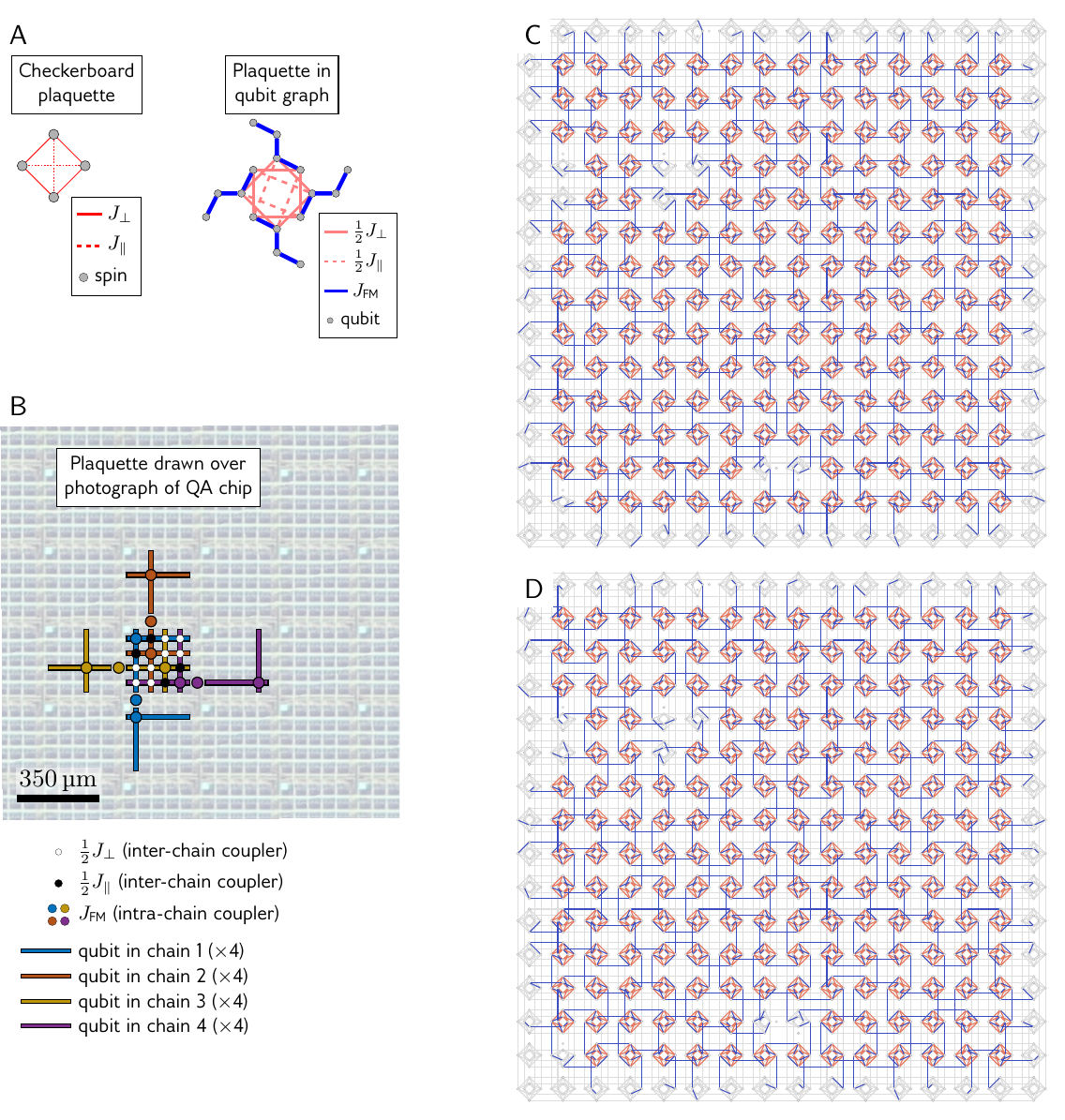}\\
FIG. S1. {\bf Embeddings of the square ice system.} {\bf A}:~Each ice vertex is represented by a plaquette in the checkerboard lattice or in the qubit graph, as pictured (cf.~ Fig.~1). {\bf B}:~Realization of an ice vertex as an Ising plaquette superimposed over a photograph of a portion of the QA processor (see text for details).  {\bf C--D}: Two possible embeddings of the $14\times 14$ lattice of ice vertices in the QA chip.  Red and blue couplings are AFM and FM respectively.  Both embeddings realize the couplings of the same Ising model using different physical qubits and couplers in the quantum annealer; 20 such embeddings are used, to suppress the effect of device variation.

\clearpage

\noindent  \includegraphics[width=\linewidth]{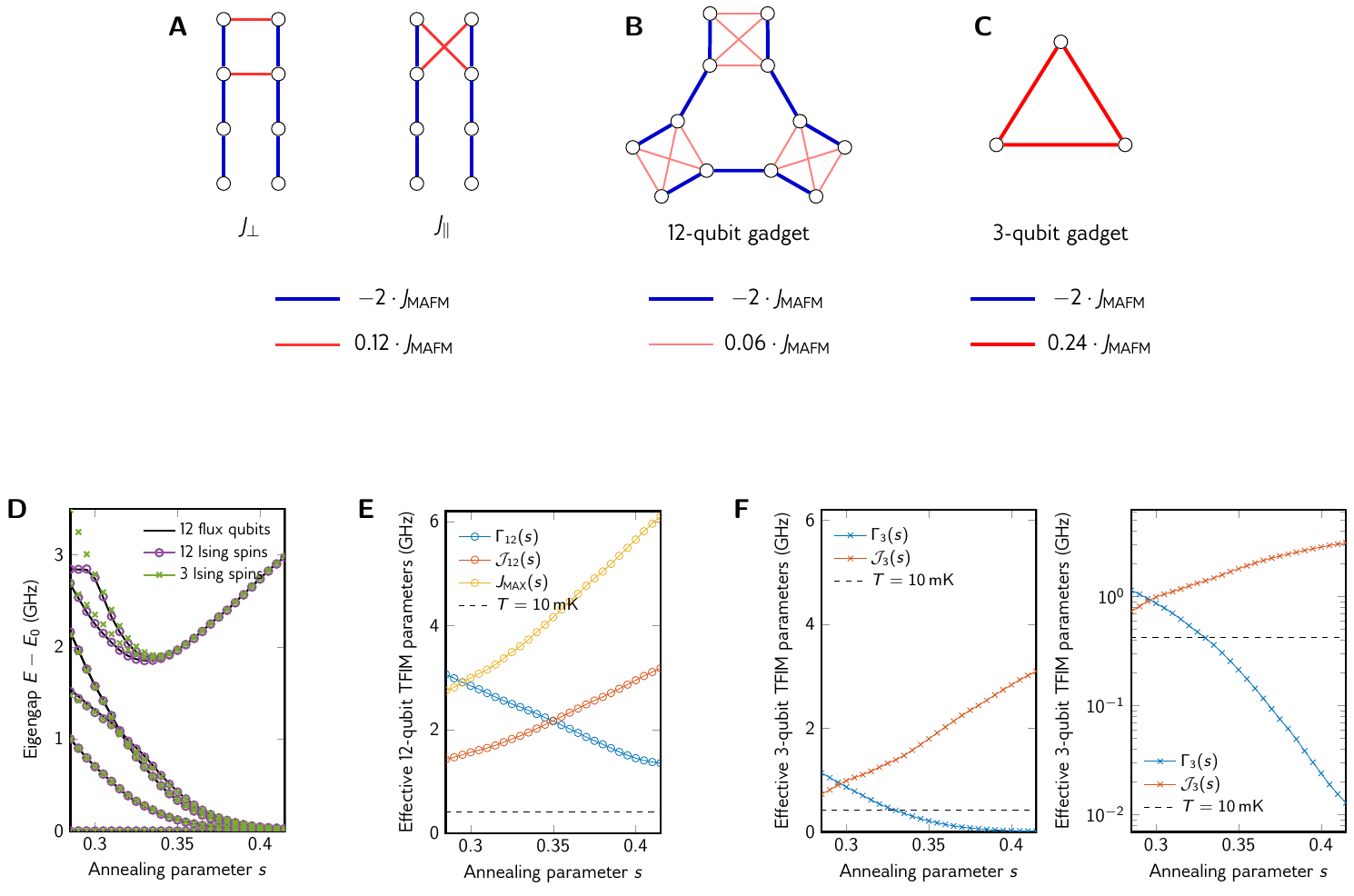}\\
  FIG. S2. {\bf Extracting effective TFIM parameters:} The Ising square ice is realized by mapping each spin in the checkerboard lattice (Fig.~1B, Fig.~3) to a ferromagnetically-coupled chain of four qubits (Fig.~1A, Fig.~S1).  In this embedding of the square ice into the ``Chimera'' qubit arrangement, $J_\perp$ couplings and $J_\parallel$ couplings are realized with slightly different geometry ({\bf A}).  To extract effective parameters of the Chimera TFIM Hamiltonian, we study a 12-variable gadget whose couplings reflect the relevant embedding geometry ({\bf B}).  Using this gadget we compute the spectrum of 12 rf-SQUID flux qubits and 12 Ising spins ({\bf D}), and determine the Ising tunneling and coupling parameters $\Gamma$ and $J_{\mathrm{MAX}}$ ({\bf E}) using a best fit on the lowest eight eigengaps.  To estimate the difference in energy scales between the embedded Chimera TFIM and the logical square ice TFIM, we also map the 12-qubit (3-chain) gadget to a 3-spin gadget ({\bf C}) and extract the effective Hamiltonian ({\bf F}).

\clearpage

 \noindent \includegraphics[width=\linewidth]{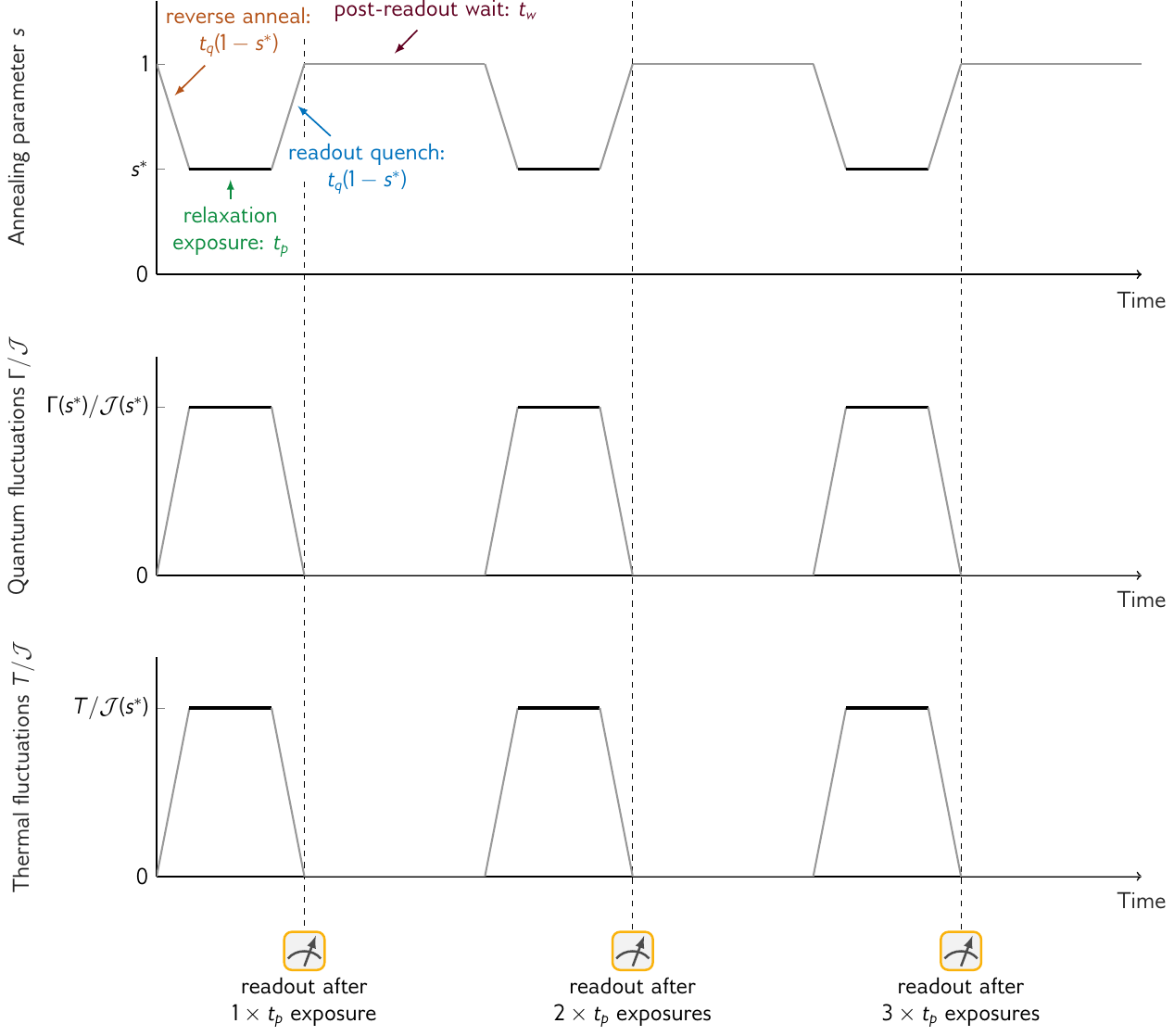}\\
 FIG. S3. {\bf Quantum annealing control:} Diagram (not to scale) of annealing parameter $s$ versus time in the QA protocol.  The system follows a cycle of 128 reverse anneals per programming.  Each reverse anneal starts with a classical state and ends with a classical state.  The system is exposed to quantum and thermal fluctuations for a fixed pause time $t_p$ before the fluctuations are quenched and the state is destructively projected to the computational ($\sigma^z$) basis for readout.  The experiments in this work use $t_q=1\ \mu s$, $t_p=256\ \mu s$ (Figs.~2, 4, S6) or $1\ \mu s$ (Figs.~3, S4, and supplemental movies), and $t_w = 0.7\ ms - t_p$.

\clearpage

\noindent \includegraphics[scale=1.3]{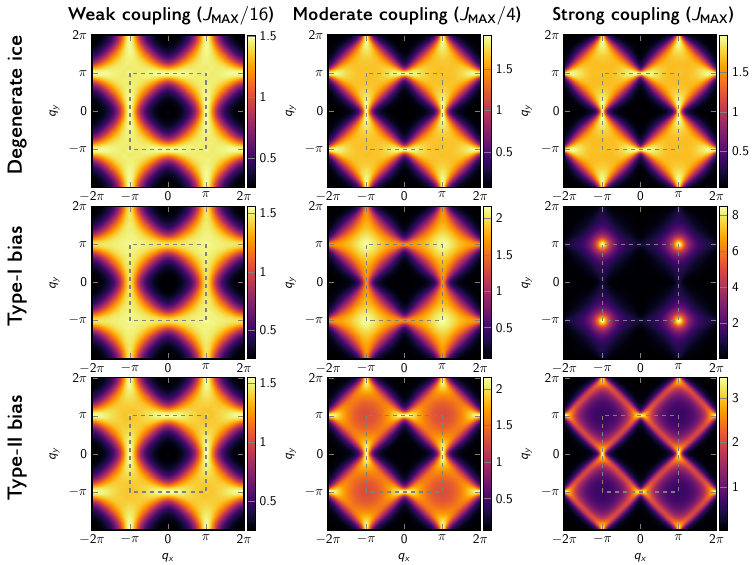}\\
FIG. S4. Real-space structure factors $S_{\mathrm{real}}(\mathbf q)$ analogous to the reciprocal-space structure factors shown in Fig.~2, with the lattice Brillouin zone shown as a dashed line.  Intensity $S_{\mathrm{real}}(\mathbf q)$ is in arbitrary units, with $x$ and $y$ components of $\mathbf q$ in reciprocal lattice units.

\clearpage

\noindent \includegraphics[scale=1.1]{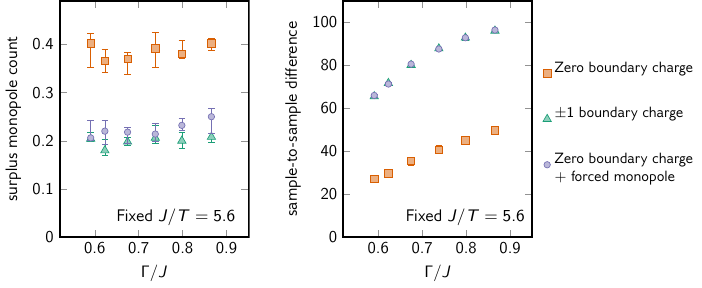}\\
FIG. S5. {\bf Effect of monopoles on mixing.}  Average surplus monopole count (excess from ground state) (left) and sample-to-sample difference (in spins) (right) for QA steps with varying values of $\Gamma/J$.  The three closed boundary conditions (Fig.~4B--D) are studied.  The two boundary cases with itinerant monopoles show fewer surplus monopoles and faster mixing, compared to the case with no monopoles in the ground state (orange squares).  As $\Gamma/J$ increases the system mixes faster without significant addition of monopole excitations, indicating dynamics driven by quantum fluctuations.  Error bars are 95\% statistical bootstrap confidence intervals on the mean across 50 QA calls, using only the last 10 of 128 samples from each call to reduce the effect of random initial states.  Boundary spins are not counted in the sample-to-sample difference.

\clearpage

  \noindent   \includegraphics[width=17cm]{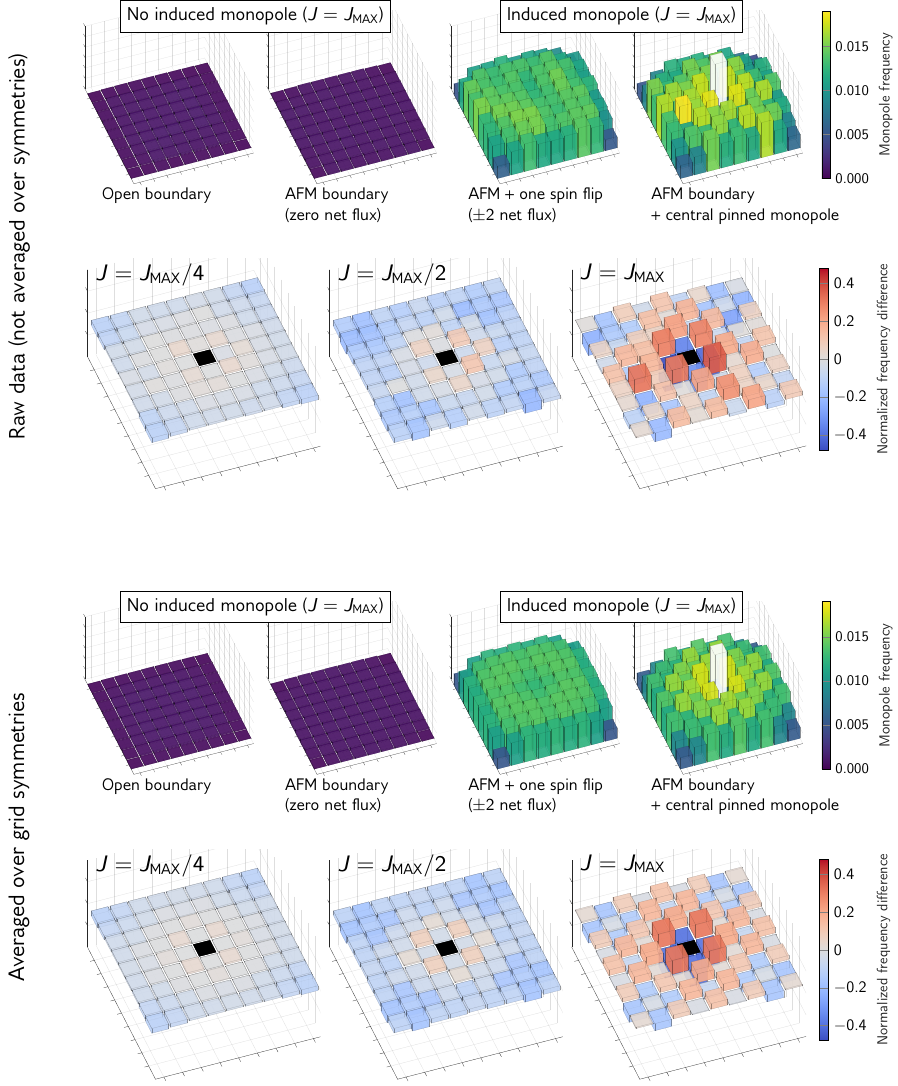}\\
FIG. S6. Monopole populations on a subgrid with no vacancies, analogous to Fig.~4.  Restricting the artificial spin ice to a $9\times 9$ subgrid allows us to study a system that has no defects.  Consequently the square grid has eight symmetries over which we can average the monopole population, further suppressing experimental variation (bottom).  This makes the entropic screening, in particular its relation to the parity of the grid distance from the forced monopole, more clear.

\clearpage

  \noindent   \includegraphics[width=11cm]{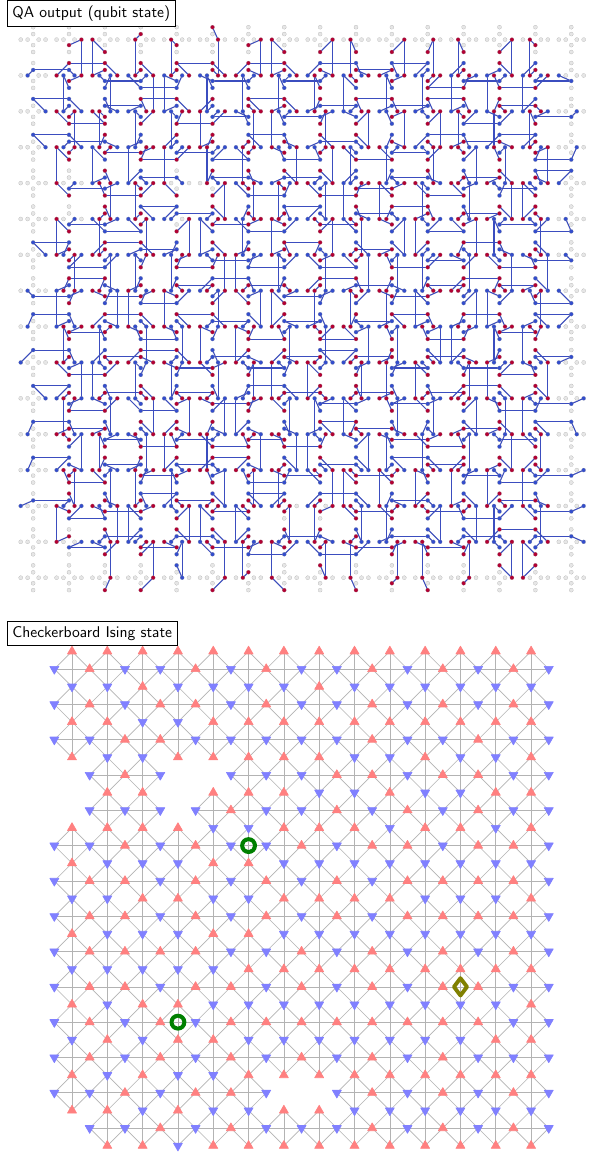}\\
  FIG. S7. Conversion of QA output to square ice state.  The QA output consists of a $\pm 1$ Ising state for 1648 of the 2048 qubits.  Four-qubit chains (see Fig.~1D) are depicted by blue lines.  Spin values of $+1$ and $-1$ are depicted by red and blue circles, respectively.  The couplings used are shown in Fig.~S1C.  A checkerboard Ising state is constructed by contracting each four-qubit chain into a single logical spin (red and blue triangles indicate $+1$ and $-1$ respectively).  This state corresponds to the dimer configuration shown in Fig.~3, top-left panel.

\clearpage
\subsection*{Movies}
Supplementary video files show examples of ice states for a chain of 128 samples separated by $1 \mu s$ exposures; Fig.~3 shows four states from such a sequence.  The examples correspond to open boundaries with varying $J_{\perp}/J_{\parallel}$ bias as in Fig.~2, and the degenerate case with varying boundary conditions as in Fig.~4.  All videos correspond to experiments with $J=J_{\mathrm{MAX}}$.\\

\noindent {\bf Movie S1.}
Degenerate ice with open boundary.  Corresponds to Fig.~2, top row, ``strong coupling'' column, and also to Fig.~4A.\\

\noindent {\bf Movie S2.}
Type-I bias with open boundary.  Corresponds to Fig.~2, middle row, ``strong coupling'' column.\\

\noindent {\bf Movie S3.}
Type-II bias with open boundary.  Corresponds to Fig.~2, bottom row, ``strong coupling'' column.\\

\noindent {\bf Movie S4.}
Degenerate ice with AFM boundary (zero flux).  Corresponds to Fig.~4B.\\

\noindent {\bf Movie S5.}
Degenerate ice with AFM boundary but with one spin flipped (nonzero flux).  Corresponds to Fig.~4C.\\

\noindent {\bf Movie S6.}
Degenerate ice with AFM boundary but with a central pinned monopole.  Corresponds to Fig.~4D.


%apsrev4-2.bst 2019-01-14 (MD) hand-edited version of apsrev4-1.bst
%Control: key (0)
%Control: author (8) initials jnrlst
%Control: editor formatted (1) identically to author
%Control: production of article title (0) allowed
%Control: page (0) single
%Control: year (1) truncated
%Control: production of eprint (0) enabled
\begin{thebibliography}{0}%
\makeatletter
\providecommand \@ifxundefined [1]{%
 \@ifx{#1\undefined}
}%
\providecommand \@ifnum [1]{%
 \ifnum #1\expandafter \@firstoftwo
 \else \expandafter \@secondoftwo
 \fi
}%
\providecommand \@ifx [1]{%
 \ifx #1\expandafter \@firstoftwo
 \else \expandafter \@secondoftwo
 \fi
}%
\providecommand \natexlab [1]{#1}%
\providecommand \enquote  [1]{``#1''}%
\providecommand \bibnamefont  [1]{#1}%
\providecommand \bibfnamefont [1]{#1}%
\providecommand \citenamefont [1]{#1}%
\providecommand \href@noop [0]{\@secondoftwo}%
\providecommand \href [0]{\begingroup \@sanitize@url \@href}%
\providecommand \@href[1]{\@@startlink{#1}\@@href}%
\providecommand \@@href[1]{\endgroup#1\@@endlink}%
\providecommand \@sanitize@url [0]{\catcode `\\12\catcode `\$12\catcode
  `\&12\catcode `\#12\catcode `\^12\catcode `\_12\catcode `\%12\relax}%
\providecommand \@@startlink[1]{}%
\providecommand \@@endlink[0]{}%
\providecommand \url  [0]{\begingroup\@sanitize@url \@url }%
\providecommand \@url [1]{\endgroup\@href {#1}{\urlprefix }}%
\providecommand \urlprefix  [0]{URL }%
\providecommand \Eprint [0]{\href }%
\providecommand \doibase [0]{https://doi.org/}%
\providecommand \selectlanguage [0]{\@gobble}%
\providecommand \bibinfo  [0]{\@secondoftwo}%
\providecommand \bibfield  [0]{\@secondoftwo}%
\providecommand \translation [1]{[#1]}%
\providecommand \BibitemOpen [0]{}%
\providecommand \bibitemStop [0]{}%
\providecommand \bibitemNoStop [0]{.\EOS\space}%
\providecommand \EOS [0]{\spacefactor3000\relax}%
\providecommand \BibitemShut  [1]{\csname bibitem#1\endcsname}%
\let\auto@bib@innerbib\@empty
%</preamble>
\end{thebibliography}%


\begin{thebibliography}{10}

\bibitem{Wang2006}
R.~F. Wang, {\it et~al.\/}, {\it Nature\/} {\bf 439}, 303 (2006).

\bibitem{tanaka2006magnetic}
M.~Tanaka, E.~Saitoh, H.~Miyajima, T.~Yamaoka, Y.~Iye, {\it Physical Review
  B\/} {\bf 73}, 052411 (2006).

\bibitem{Ramirez1999}
A.~P. {Ramirez}, A.~{Hayashi}, R.~J. {Cava}, R.~{Siddharthan}, B.~S. {Shastry},
  {\it Nature\/} {\bf 399}, 333 (1999).

\bibitem{Bramwell2001}
S.~T. Bramwell, M.~J. Gingras, {\it Science\/} {\bf 294}, 1495 (2001).

\bibitem{nisoli2017deliberate}
C.~Nisoli, V.~Kapaklis, P.~Schiffer, {\it Nature Physics\/} {\bf 13}, 200
  (2017).

\bibitem{Castelnovo2008}
C.~Castelnovo, R.~Moessner, S.~L. Sondhi, {\it Nature\/} {\bf 451}, 42 (2008).

\bibitem{ryzhkin2005magnetic2}
I.~Ryzhkin, {\it Journal of Experimental and Theoretical Physics\/} {\bf 101},
  481 (2005).

\bibitem{perrin2016extensive}
Y.~Perrin, B.~Canals, N.~Rougemaille, {\it Nature\/} {\bf 540}, 410 (2016).

\bibitem{farhan2019emergent}
A.~Farhan, {\it et~al.\/}, {\it Science Advances\/} {\bf 5}, eaav6380 (2019).

\bibitem{Nisoli2013colloquium}
C.~Nisoli, R.~Moessner, P.~Schiffer, {\it Reviews of Modern Physics\/} {\bf
  85}, 1473 (2013).

\bibitem{skjaervo2019advances}
S.~H. Skj{\ae}rv{\o}, C.~H. Marrows, R.~L. Stamps, L.~J. Heyderman, {\it Nature
  Reviews Physics\/} pp. 1--16 (2019).

\bibitem{ortiz2019colloquium}
A.~Ortiz-Ambriz, C.~Nisoli, C.~Reichhardt, C.~J. Reichhardt, P.~Tierno, {\it
  Reviews of Modern Physics\/} {\bf 91}, 041003 (2019).

\bibitem{meeussen2020topological}
A.~S. Meeussen, E.~C. O{\u{g}}uz, Y.~Shokef, M.~van Hecke, {\it Nature
  Physics\/} pp. 1--5 (2020).

\bibitem{Harris2010a}
R.~Harris, {\it et~al.\/}, {\it Physical Review B\/} {\bf 81}, 1 (2010).

\bibitem{Henry2014}
L.-P. Henry, T.~Roscilde, {\it Physical Review Letters\/} {\bf 113}, 1 (2014).

\bibitem{Chamon2019}
C.~Chamon, D.~Green, Z.-C. Yang, {\it Phys. Rev. Lett.\/} {\bf 125}, 067203
  (2020).

\bibitem{Bunyk2014}
P.~I. Bunyk, {\it et~al.\/}, {\it IEEE Transactions on Applied
  Superconductivity\/} {\bf 24}, 1 (2014).

\bibitem{sm} See Supplementary Materials.

\bibitem{gingras2014quantum}
M.~J. Gingras, P.~A. McClarty, {\it Reports on Progress in Physics\/} {\bf 77}
  (2014).

\bibitem{Harris2018}
R.~Harris, {\it et~al.\/}, {\it Science\/} {\bf 361}, 162 (2018).

\bibitem{King2018}
A.~D. King, {\it et~al.\/}, {\it Nature\/} {\bf 560}, 456 (2018).

\bibitem{Weinberg2019}
P.~Weinberg, {\it et~al.\/}, {\it Physical Review Letters\/} {\bf 124}, 090502
  (2020).

\bibitem{King2019}
A.~D. King, {\it et~al.\/}, {\it Nature Communications\/} {\bf 12}, 1113 (2021).

\bibitem{lieb1967residual}
E.~H. Lieb, {\it Physical Review\/} {\bf 162}, 162 (1967).

\bibitem{Morgan2011}
J.~P. Morgan, A.~Stein, S.~Langridge, C.~H. Marrows, {\it Nature Physics\/}
  {\bf 7}, 75 (2011).

\bibitem{henley2010coulomb}
C.~L. Henley, {\it Annu. Rev. Condens. Matter Phys.\/} {\bf 1}, 179 (2010).

\bibitem{Porro2013}
J.~M. Porro, A.~Bedoya-Pinto, A.~Berger, P.~Vavassori, {\it New Journal of
  Physics\/} {\bf 15} (2013).

\bibitem{farhan2013direct}
A.~Farhan, {\it et~al.\/}, {\it Physical Review Letters\/} {\bf 111}, 057204
  (2013).

\bibitem{kapaklis2014thermal}
V.~Kapaklis, {\it et~al.\/}, {\it Nature Nanotechnology\/} {\bf 9}, 514 (2014).

\bibitem{ortiz2016engineering}
A.~Ortiz-Ambriz, P.~Tierno, {\it Nature Communications\/} {\bf 7} (2016).

\bibitem{nisoli2020field}
C.~Nisoli, {\it Physical Review B\/} {\bf 102}, 220401 (2020).

\bibitem{morrison2013unhappy}
M.~J. Morrison, T.~R. Nelson, C.~Nisoli, {\it New Journal of Physics\/} {\bf
  15}, 045009 (2013).

\bibitem{chern2014realizing}
G.-W. Chern, C.~Reichhardt, C.~Nisoli, {\it Applied Physics Letters\/} {\bf
  104}, 013101 (2014).

\bibitem{stern2019quantum}
M.~Stern, C.~Castelnovo, R.~Moessner, V.~Oganesyan, S.~Gopalakrishnan, {\it
  arXiv:1911.05742\/}  (2019).

\bibitem{data}
A.~D.~King, C.~Nisoli, E.~D.~Dahl, G.~Poulin-Lamarre, and A.~Lopez-Bezanilla, {\it Zenodo} (2021). Data set.  http://doi.org/10.5281/zenodo.4776361.


\bibitem{Johnson2011}
M.~W.~Johnson, {\it et~al.\/}, {\it Nature\/} {\bf 473}, 194 (2011).

\bibitem{Kairys2020}
P. Kairys, {\it et~al.\/}, {\it PRX Quantum\/} {\bf 1}, 020320 (2020).


\end{thebibliography}
\end{document}